# Zero field entanglement in dipolar coupling spin system at negative temperature

22 September 2013

*Gregory B. Furman , Victor M. Meerovich , and Vladimir L. Sokolovsky*
*Physics Department, Ben Gurion University of the Negev, Beer Sheva, 84105 Israel*

A dipolar coupled spin system can achieve internal thermodynamic equilibrium states at negative absolute temperature. We study analytically and numerically the temperature dependence of the concurrence in a dipolar coupled spin-1/2 system in both non-zero and zero fields and show that, at negative temperatures, entangled states can exist even in zero magnetic field.

## Introduction

Quantum entanglement in spin systems has been intensively studied during the last two decades [1-3]. Most of these studies investigated spin-spin interacting systems located in an external magnetic field [2-8]. Various of initial conditions and the state of a quantum system have been considered in order to obtain the maximally entangled states [9, 10]. It was stated that a magnetic field is required to create entangled state in a dipolar coupled spin system [9, 11-13]. However, when this magnetic field is much larger than the local field characterizing the spin-spin interaction, entanglement in the spin system disappears [9, 11-13].

Spin-spin interactions play the decisive role in the formation of entangled states. Spin systems in thermal equilibrium became entangled if the interaction energy $\varepsilon$ between spins is larger than the thermal energy due to spin coupling to the environment, $\frac{\varepsilon}{kT} > 1$ [5-8, 9], where $k$ is the Boltzmann constant, $T$ is the temperature. For example, entanglement in a dipolar coupled spin system in equilibrium is achieved by the application of an external magnetic field in which the Zeeman interaction energy is the order of or even less than the dipolar interaction energy, $\omega_0 \leq \omega_d$ [6, 7, 14]. For proton spin systems in solids, the dipolar energies are in the range of $\omega_d = 10 \div 100$ kHz. This results in the temperature of the entanglement appearance of $T \sim \frac{\omega_d}{k_B} \sim 0.4 \div 4 \mu K$ [5-8] and it is confirmed by direct calculations [5-8].

In this paper, we show that it is possible to create the conditions under which entangled states exist without an external magnetic field in dipolar coupled spin systems.

The structure of the paper is as follows: in the next section we investigate the temperature dependence of the pairwise concurrence for a spin system located in a magnetic field, both at positive and negative temperatures. In Sec. III we present zero field entanglement. Discussion of the results is given in the final section. We consider the structures of actually existed molecules with various numbers of the spins in the form of linear chains and circles.

## Equilibrium states of dipolar coupling spins at negative temperature

According to the basic hypothesis of the spin-temperature theory, the evolution of an isolated spin system of a large number of dipolar coupled spins in an external magnetic field leads to state of thermal equilibrium described by the following density matrix [15, 16]:

$$\rho = \frac{\exp\left(-\frac{H}{kT}\right)}{Tr\left\{\exp\left(-\frac{H}{kT}\right)\right\}}, \quad (1)$$

where $T$ is the spin temperature. The form of the Hamiltonian $H$ depends on the type of the spin-spin interactions and for a system of identical dipole-dipole coupling spins 1/2 in the external field the Hamiltonian can be represented as a sum of the Zeeman

$$H_Z = \omega_0 \sum_{k=1}^{N} I_k^z \quad (2)$$

and the spin-spin coupling

$$H_{dd} = -\gamma^2 \hbar \sum_{mn} \left( \frac{3}{r_{mn}^5} (I_m r_{mn})(I_n r_{mn}) - \frac{1}{r_{mn}^3} I_m I_n \right) \quad (3)$$

terms. Here $\omega_0 = \gamma H_0$, $\gamma$ is the gyromagnetic ratio, $I_m$ is the angular momentum operator of the $m$-th spin ($m = 1, 2...N$), $r_{mn}$ is the radius vector from the $m$-th to $n$-th spins with the spherical coordinates $r_{mn}$, $\theta_{mn}$ and $\varphi_{mn}$, $\theta_{mn}$ is the angle between the radius vector $r_{mn}$ and direction of the external magnetic field.

The equilibrium distribution (1) of populations among the energy levels is provided due to the presence of so-called flip-flop terms in Hamiltonian (3) [15, 16]. The time required for achieving this distribution is the order of the time of the free-induction decay $T_2 \sim \omega_d^{-1}$ [15, 16]. We note that the considered spin systems satisfy the requirements for achieving an equilibrium states at a negative temperature [17-20]: (a) the system has a finite number of energy levels, (b) dipole-dipole interactions provide establishing internal thermodynamic equilibrium and (c) during time $T_2$ the spin system can be considered as thermally isolated from surroundings. Equilibrium states at a negative temperature has been experimentally achieved in a spin system of LiF crystal with dipole-dipole interactions [18], in an equilibrated Mott insulator of rubidium atoms [21] and in cold $^{39}K$ atoms of optical lattices [22]. Recently, it was demonstrated the appearance of entangled states in quantum systems at negative temperatures [23]. Developing this approach, let us investigate in details the temperature and field behavior of entanglement in the considered system over the entire temperature range.

## Measure of entanglement

We will characterize the entangled states by the concurrence between two, the $m-$th and $n-$th, spins [24]. For density matrices, the concurrence is defined as

$$C_{mn}(\rho_{mn}) = \max(0, F_{mn}) \quad (4)$$

where $\rho_{mn}$ is the reduced density matrix. For the $m$-th and $n$-th spins, the reduced density matrix $\rho_{mn}$ is defined as $\rho_{mn} = Tr_{mn}(\rho)$ where $Tr_{mn}(...)$ denotes the trace over the degrees of freedom for all spins except the $m$-th and $n$-th spins.

$$F_{mn} = \lambda_1^{mn} - \lambda_2^{mn} - \lambda_3^{mn} - \lambda_4^{mn} \quad (5)$$

$\lambda_i$ are the eigenvalues, in decreasing order, of the operator

$$R_{mn} = \sqrt{\sqrt{\rho_{mn}} \tilde{\rho}_{mn} \sqrt{\rho_{mn}}} \qquad (6)$$

and

$$\tilde{\rho}_{mn} = (\sigma_y \otimes \sigma_y) \rho^*_{mn} (\sigma_y \otimes \sigma_y). \qquad (7)$$

In Eq. (21) $\tilde{\rho}_{mn}$ is the complex conjugation of the reduced density matrix $\rho_{mn}$ and $\sigma_y$ is the Pauli matrix $\sigma_y = \begin{pmatrix} 0 & -i \\ i & 0 \end{pmatrix}$. For maximally entangled states, the concurrence is $C_{mn} = 1$ while for separable states $C_{mn} = 0$.

To clarify the peculiarities of the entangled states at a negative temperature, let us consider a system of two spins. An analytical expression can be obtained for a two-spin system at $\theta_{mn} = \frac{\pi}{2}$ and $\varphi_{mn} = 0$ $(j = 1, k = 2)$ [12]. The eigenvalues $\lambda_i$ (we omit the indices $m, n$) of the operator (6) are

$$\lambda_1 = \frac{1}{Z} e^{\frac{\beta}{4}}, \quad \lambda_2 = \frac{1}{Z} e^{\frac{3\beta}{4}}, \quad \lambda_{3,4} = \frac{1}{Z} \sqrt{2B\left(B \pm 3\sinh\frac{A\beta}{4}\right) - A^2}, \qquad (8)$$

where $Z = 2A\left[e^{\frac{\beta}{2}} \cosh\frac{\beta}{4} + \cosh\frac{A\beta}{4}\right]$, $A = \sqrt{9 + 16\alpha^2}$, $B = \sqrt{9\cosh\frac{A\beta}{4} + 16\alpha^2}$, $\beta = \frac{\gamma^2 \hbar}{r_{12}^3 kT}$ and $\alpha = \frac{\omega_0 r_{12}^3}{\gamma^2 \hbar}$ are the dimensionless inverse temperature and magnetic field strength, respectively.

The dependences of the eigenvalues and concurrence on the inverse temperature are asymmetrical relative $\beta = 0$ (Fig. 1). For the system in the magnetic field the maximal eigenvalues differ for positive and negative temperatures: at $T > 0$ $\lambda_4$ is maximal while at $T < 0$ $\lambda_3$ is maximal (Fig. 1a). At $|\beta| \gg 1$ the concurrence is determined only by these eigenvalues and $\lambda_3(\beta) = \lambda_4(-\beta)$, hence, $C(\beta) = C(-\beta)$ (Fig. 1b).

Figure 2 presents the concurrence as a function of the magnetic field over the entire range of negative and positive temperatures. The concurrence increases with $|\beta|$ and at $|\beta| \to \infty$ achieves a limit which depends on the magnetic field. On the phase plane $\beta - \alpha$ (Figure 3) the areas of the entangled states at positive and negative inverse temperatures are isolated from each other by the area of the separated state. The entangled states are observed at $|\beta|$ higher than a critical value $\beta_{cr}$ which is a function of the applied magnetic field $\alpha$ (Figures 2 and 3). The dependence of $\beta_{cr}$ on the magnetic field is different for negative and positive values of $\beta$: at negative values $\beta_{cr}$ decreases approximately linearly (slop ~ 0.1 ) with increasing $\alpha$, whereas at positive values, this dependence is close to $\frac{1}{\alpha}$. In both cases, at $\alpha \gg 1$ the concurrence changes as $\frac{1}{\sqrt{\alpha}}$. The obtained results show that in the negative temperature region entanglement exists even at zero magnetic field.

## Eigenfunction and concurrence of identical spins at zero field

To clarify the peculiarities of the entangled states at zero magnetic field, we first consider a system of two identical spins $1/2$. One of the ways to reach a zero field state [25-29] of a spin system is using the adiabatic demagnetization method [15, 16]. At zero field, we choose the quantization $z$-axis along the vector $r_{12}$ from the first to the second spin [30]. The Hamiltonian of the dipolar interaction takes very simple form which can be obtained from (3) at $\theta = 0$ and $\varphi = 0$

$$H_{dd} = D\left[I_1^z I_2^z - \frac{1}{4}\left(I_1^+ I_2^- + I_1^- I_2^+\right)\right], \quad (9)$$

where $D = \frac{2\gamma^2}{r_{12}^3}$. Diagonalization of the operator (10) on the basis of eigenfunctions of the operator $I^z = I_1^z + I_2^z$ dives the following eigenvalues

$$\xi_1 = -\frac{D}{2}, \; \xi_2 = 0, \; \xi_3 = D, \; \xi_4 = -\frac{D}{2}, \quad (10)$$

and the eigenvectors which correspond to the following system states

$$|\uparrow\uparrow\rangle_1 = \begin{Bmatrix} 1 \\ 0 \\ 0 \\ 0 \end{Bmatrix}; \; |\uparrow\downarrow\rangle - |\downarrow\uparrow\rangle_2 = \begin{Bmatrix} 0 \\ -\frac{1}{2} \\ \frac{1}{2} \\ 0 \end{Bmatrix}; \; |\uparrow\downarrow\rangle + |\downarrow\uparrow\rangle_3 = \begin{Bmatrix} 0 \\ \frac{1}{2} \\ \frac{1}{2} \\ 0 \end{Bmatrix}; \; |\downarrow\downarrow\rangle_4 = \begin{Bmatrix} 0 \\ 0 \\ 0 \\ 1 \end{Bmatrix}.$$

(11)

where $\uparrow$ and $\downarrow$ note the up and down spin states, respectively.

An equilibrium state is achieved in the time, and is described by the density matrix (1) with the Hamiltonian (10). In the chosen basis the density matrix takes the following form

$$\rho_0 = \frac{1}{Z_0}\begin{pmatrix} e^\beta & 0 & 0 & 0 \\ 0 & \cosh\frac{\beta}{2} & -\sinh\frac{\beta}{2} & 0 \\ 0 & -\sinh\frac{\beta}{2} & \cosh\frac{\beta}{2} & 0 \\ 0 & 0 & 0 & e^\beta \end{pmatrix} \quad (12)$$

where $Z_0 = 2\left(\cosh\frac{\beta}{2} + e^\beta\right)$ is the partition function.

The eigenvalues of the operator $R_{12}$ can be obtain directly by using (6) or by using the zero field limit in Eq. (8). The result is

$$\lambda_1 = \frac{e^\beta}{Z_0}, \; \lambda_2 = \frac{e^{-\frac{\beta}{2}}}{Z_0}, \; \lambda_3 = \frac{e^\beta}{Z_0}, \lambda_4 = \frac{e^{\frac{\beta}{2}}}{Z_0}. \quad (13)$$

At a positive temperature, $\beta > 0$, two maximal eigenvalues, $\lambda_1$ and $\lambda_3$, are equal that leads to $K_{12} < 0$, which implies the impossibility of creation of an entangled state of the dipolar coupling spins in zero magnetic field at a positive temperature.

At a negative temperature, $\beta < 0$, $\lambda_2$ is the maximal eigenvalue, and the entangled state is observed at $\beta < \beta_{cr}$. $\beta_{cr} \approx -0.839$ as determined from the condition $K_{12} = 0$. The black curve (Fig. 4) presents the concurrence for a two-spin system as a function of the inverse temperature at zero magnetic field. The concurrence increases with decreasing temperature and, at $\beta \to -\infty$, it is reached the maximum value $C = 1$.

The pairwise concurrences for 6-, and 8- spin systems, circles and chains, demonstrate similar properties: there are negative critical temperatures below which systems are entangled and no entangled states at a positive temperature (Fig. 4). The considered systems simulate structures of real compounds which can be used in experiments: e.g. circle of six spins - proton spins of benzene molecule $C_6H_6$; circle of eighth spins - cyclooctatetraene dipotassium with chemical formula $C_8H_8K_2$ [31]; chains - proton and fluorine chains in calcium hydroxyapatite $Ca_5(OH)(PO_4)_3$ or in calcium fluorapatite $Ca_5F(PO_4)_3$ [32]. The maximal concurrence is achieved in a two-spin system (Fig. 4, solid black curve). In systems with higher spin number the pairwise concurrence is lower, that can be explained by entanglement of these spins with other spins of the system. For example, the pairwise concurrence between the first and the second spins equals the concurrence between the first and the last spins in a circular system. In a linear system the location of a spin influences its pairwise concurrence with other spins (e.g. compare shot-dashed magenta and dashed red curves, Fig. 4). The concurrence between the first and second spins increases monotonically with decreasing temperature while the concurrence between the second and third spins has the maximum at $\beta = -2.3$ and decreases to $0.08$ at $\beta \to \infty$.

The critical temperatures slightly depend on the spin number in the system. To estimate the critical temperature, below which entanglement exists, let us consider hydrogen with $\gamma = 4.2577 \frac{kHz}{G}$ and the dipolar interaction energy of the order of $10\ kHz$ (in frequency units) [15], which leads to the critical temperature as - $7\ \mu K$.

## Discussion and Conclusion

We obtained that the thermal dependence of the concurrence is significantly different for positive and negative temperatures. This difference is most clearly manifested in zero field where at positive temperatures entanglement completely disappears while at negative temperatures the concurrence reaches very high values (Fig. 4). The deference between negative and positive temperatures lies only in various populations of energy levels: at a positive temperature, $T > 0$, the equilibrium state is characterized by a greater population of lower energy levels than higher ones, while at negative temperature, $T < 0$, the population of the higher levels is greater than that of the lower levels [17]. Therefore, negative temperature means that the most spins are in a high-energy state, with a few in a low-energy state, so that the exponent in the density matrix (1) rises at positive eigenvalues of the Hamiltonian $H$ rather than falling and vice versa. In particular, from (1) at $\beta \to \infty$ we obtain

$$\rho_0 = \frac{1}{Z_0}\begin{pmatrix} 1 & 0 & 0 & 0 \\ 0 & 0 & 0 & 0 \\ 0 & 0 & 0 & 0 \\ 0 & 0 & 0 & 1 \end{pmatrix} \quad (14)$$

while at $\beta \to -\infty$ we have

$$\rho_0 = \frac{1}{Z_0} \begin{pmatrix} 0 & 0 & 0 & 0 \\ 0 & 1 & 1 & 0 \\ 0 & 1 & 1 & 0 \\ 0 & 0 & 0 & 0 \end{pmatrix} \quad (15)$$

It is known [1-3] that matrix (14) corresponds to a separated state with $C=0$, while matrix (15) -- to a maximally entangled state with $C=1$.

An important question remains about methods for reaching negative temperatures. The method depends on the quantum system. As noted above, quantum system which is isolated from the environment and possesses an upper bound in energy, so quantum objects that make up the system could pile up in high-energy rather than low-energy states, is an ideal system to create negative temperature states. Of course the method of achieving negative temperatures depends on the quantum system. For example states with negative temperature can be attained in practice in a paramagnetic system of nuclear spins in a crystal [17, 18]. The crystal will be magnetized in a strong magnetic field, and then the direction of the field will be quickly reversed. The system is thus in a non-equilibrium state. During a time of the order of $T_2$, the system reaches an equilibrium state. If the field is then adiabatically removed, the system remains in the equilibrium state, which will have a negative temperature. The subsequent exchange of energy between the spin system and the lattice, whereby their temperatures are equalized, takes place in a time of the order of $T_1 \gg T_2$.

*In conclusion*, the investigation of zero-field entanglement in a spin system under the thermodynamic equilibrium conditions showed that the entangled state can be achieved only at negative temperatures and never at positive temperatures. Thus, entangled states in dipolar coupled spin systems can be created under zero field conditions.

Figures

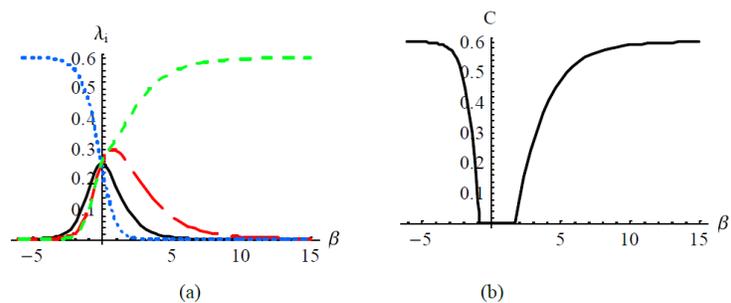

Fig. 1 The eigenvalues $\lambda_i$ (a) and concurrence (b) vs. temperature at $\alpha = 1$. Black solid curve $\lambda_1$, red dashed -- $\lambda_2$, blue dotted - $\lambda_3$, green short-dashed - $\lambda_4$.

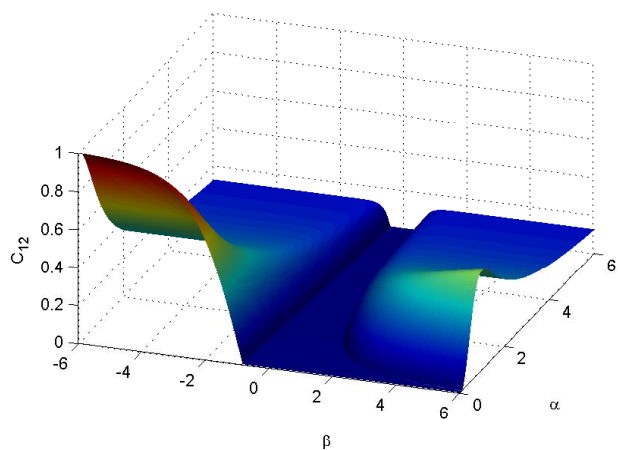

Fig. 2 Concurrence of dipole-dipole coupling spins vs. temperature and magnetic field.

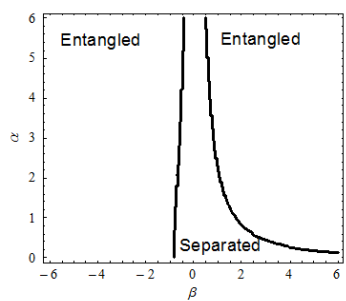

Fig. 3 Phase diagram on the plane $\beta - \alpha$.

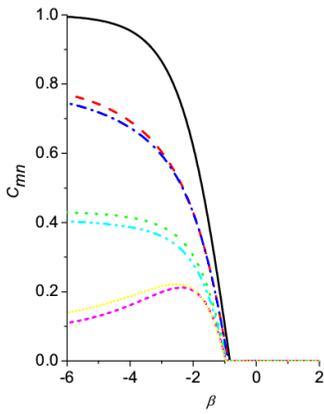

Fig. 4. Zero-field concurrence, $C_{mn}$ as a function of the inverse temperature $\beta$ : solid black curve -- $C_{12}$ in a two-spin system; dashed red -- $C_{12}$ in a six chain; dot-dashed blue -- $C_{12}$ in a eight spin chain; dotted green -- $C_{12}$ in a six spin circle; dot-dot dashed -- $C_{12}$ in a eight spin circle; shot-dashed magenta -- $C_{23}$ in a six spin chain; shot-dotted yellow -- $C_{23}$ in a eight spin chain.